# IDeF-X HD: a CMOS ASIC for the readout of Cd(Zn)Te Detectors for space-borne applications

O. Gevin[1], F. Lugiez[1], A. Michalowska[1], A. Meuris[2], O. Limousin[2], E. Delagnes[1], O. Lemaire[1], F. Pinsard[2]

[1] IRFU, CEA, Université Paris-Saclay, F-91191, Gif-sur-Yvette, France
[2] AIM, CEA, CNRS, Université Paris-Saclay, Université Paris Diderot, Sorbonne Paris Cité, F-91191 Gif-sur-Yvette, France

*Abstract*—IDeF-X HD is a 32-channel analog front-end with self-triggering capability optimized for the readout of 16 × 16 pixels CdTe or CdZnTe pixelated detectors to build low power micro gamma camera. IDeF-X HD has been designed in the standard AMS CMOS 0.35 µm process technology. Its power consumption is 800 µW per channel. The dynamic range of the ASIC can be extended to 1.1 MeV thanks to the in-channel adjustable gain stage. When no detector is connected to the chip and without input current, a 33 electrons rms ENC level is achieved after shaping with 10.7 µs peak time. Spectroscopy measurements have been performed with CdTe Schottky detectors. We measured an energy resolution of 4.2 keV FWHM at 667 keV ($^{137}$Cs) on a mono-pixel configuration. Meanwhile, we also measured 562 eV and 666 eV FWHM at 14 keV and 60 keV respectively ($^{241}$Am) with a 256 small pixel array and a low detection threshold of 1.2 keV. Since IDeF-X HD is intended for space-borne applications in astrophysics, we evaluated its radiation tolerance and its sensitivity to single event effects. We demonstrated that the ASIC remained fully functional without significant degradation of its performances after 200 krad and that no single event latch-up was detected putting the Linear Energy Transfer threshold above 110 MeV/(mg/cm$^2$). Good noise performance and radiation tolerance make the chip well suited for X-rays energy discrimination and high-energy resolution. The chip is space qualified and flies on board the Solar Orbiter ESA mission launched in 2020.

## 1 Introduction

Thanks to its attractive atomic properties (high Z, high density and wide band gap) and permanent progress in electrode lithography and crystal growth technologies, Cd(Zn)Te has become an inevitable material for building high energy resolution spectro-imaging systems operating at room temperature in the hard X ray-soft Gamma Ray range. In most cases, the pixelated or stripped Cd(Zn)Te detector is connected at least to one readout *Application Specific Integrated Circuit* (ASIC). The architecture and the performances of the readout ASICs result from the considered application that generally drives the energy range, the count rate, the sensitive material (CdTe or CdZnTe) and thus the main electric features of the detectors (input capacitance, leakage current). Medical applications often need high count rates (>1 million counts per second), low energy range (below 200 keV) and very high spatial resolution [1]-[4] whereas space astrophysics applications usually need low or moderate count rates, very high energy resolution (typically <1keV FWHM at 60 keV), high spatial resolution from the X to the hard X-rays, paving the way to high sensitivity instruments capable for instance, to resolve faint point sources supposed to populate the cosmic X- (and gamma-) ray background (CXB) [5]-[7].

For few years, our group has been developing a family of ASICs for space applications, named IDeF-X standing for Imaging Detector Front-end [8]-[11]. IDeF-X HD is the latest space qualified version of the IDeF-X family. It has been optimized for the readout of 16 × 16 pixels CdTe or CdZnTe pixelated detectors, and hence is perfectly suited to build a new low power Caliste 256 pixels module [12]. This micro gamma camera called Caliste HD [13] is the elementary unit of the MACSI (Modular Assembly of Caliste Spectro Imager) camera [14], a 2048-pixel 8 cm² gamma camera designed with 8 identical Caliste HD modules. IDeF-X HD is also the readout ASIC of Caliste-SO, a variant of a pixel detector suited to the requirements of STIX, an X-ray imaging spectrometer for solar flares observation on-board the Solar Orbiter mission [7, 15, 21] in flight since February 2020.

The paper is structured as follows:
Section II focuses on the circuit design from a global description of the ASIC features to a more detailed description of each stage of a channel. Most significant measurement results on new functions are also reported. Section III is devoted to the whole chip measurement results including gain, linearity and noise. Section IV describes the spectroscopy measurement results when IDeF-X HD is connected to different kind of detectors. Section V summarizes the measurements performed to characterize the radiation tolerance of the chip emphasizing total ionizing dose effects and latch-up characterization.

# 2 Circuit design

## *2.1 Architecture overview*

IDeF-X HD is derived from the IDeF-X SX0 [16] prototype chip to which additional functionalities have been added.

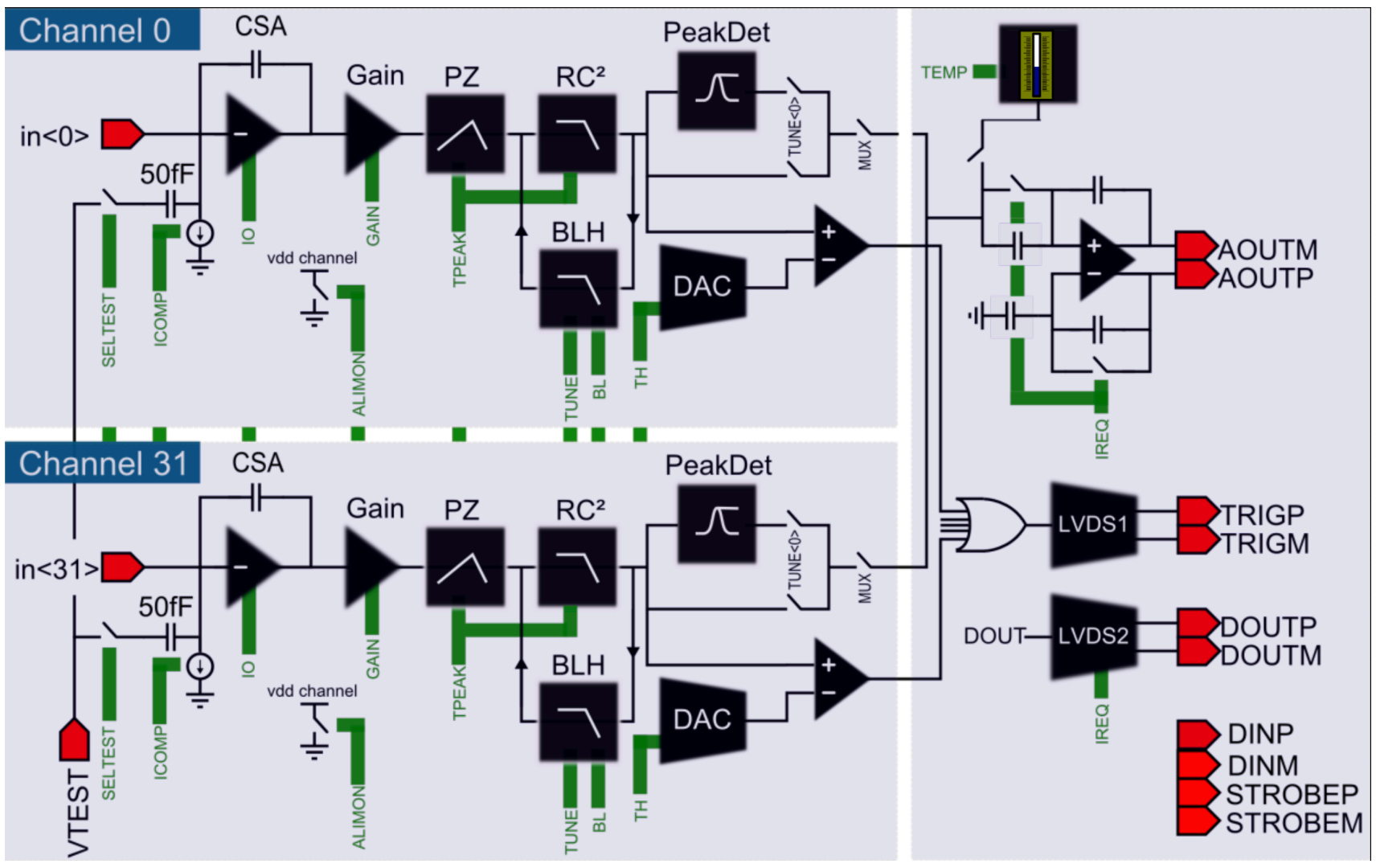

Fig. 1. Schematic of the IDeF-X HD ASIC: 32 low power channels for energy measurement.

IDeF-X HD is a 32-channel analog front-end with self-triggering capability. The architecture of the analog channel includes a Charge Sensitive Amplifier (CSA) with a continuous reset system [17], a variable gain stage (Gain), a Pole-Zero cancellation stage (PZ), a variable shaping time second order low pass filter (RC² Filter), a Baseline Holder (BLH), a peak detector (Peak Det), and a discriminator. To reduce the influence of the input charge on the Equivalent Noise Charge (ENC), we also integrated a Non-Stationary Noise Suppressor (NSNS) in the charge preamplifier stage. Each channel can be switched off (ALIMON) to reduce the total power consumption when using only few channels of the whole chip.

The energy thresholds can be individually set in each channel with the in-channel 6-bit low power DAC. The channel has been optimized to reduce the power consumption down to 800 µW per channel. The dynamic range of the ASIC can be extended to nearly 1 MeV (CdTe) without significant deterioration of the noise performances thanks to the in-channel variable gain stage. In addition, the chip includes an absolute temperature sensor capable of measuring the temperature of the Caliste module without prior thermal balance. This sensor is very useful for the full detection chain calibration. The general architecture of the chip is depicted in Fig. 1 and the layout of the chip with 32 parallel channels in Fig. 2.

In order to reduce the number of input/output signals in the Caliste module, we developed a new multi-ASIC digital serial link for both the chip configuration and readout: When using several ASICs in parallel, an address can be attached to each IDeF-X HD chip and all the input/output signals can then be shared between ASICs. To reduce potential parasitic coupling between analog inputs and digital signals, all digital signals are LVDS type with programmable current. All the digital blocks have been designed with our Single Event Latch-up (SEL) hardened library.

Every tunable parameter (gain, peak time, thresholds, test modes, CSA current…) may be set with this new serial link.

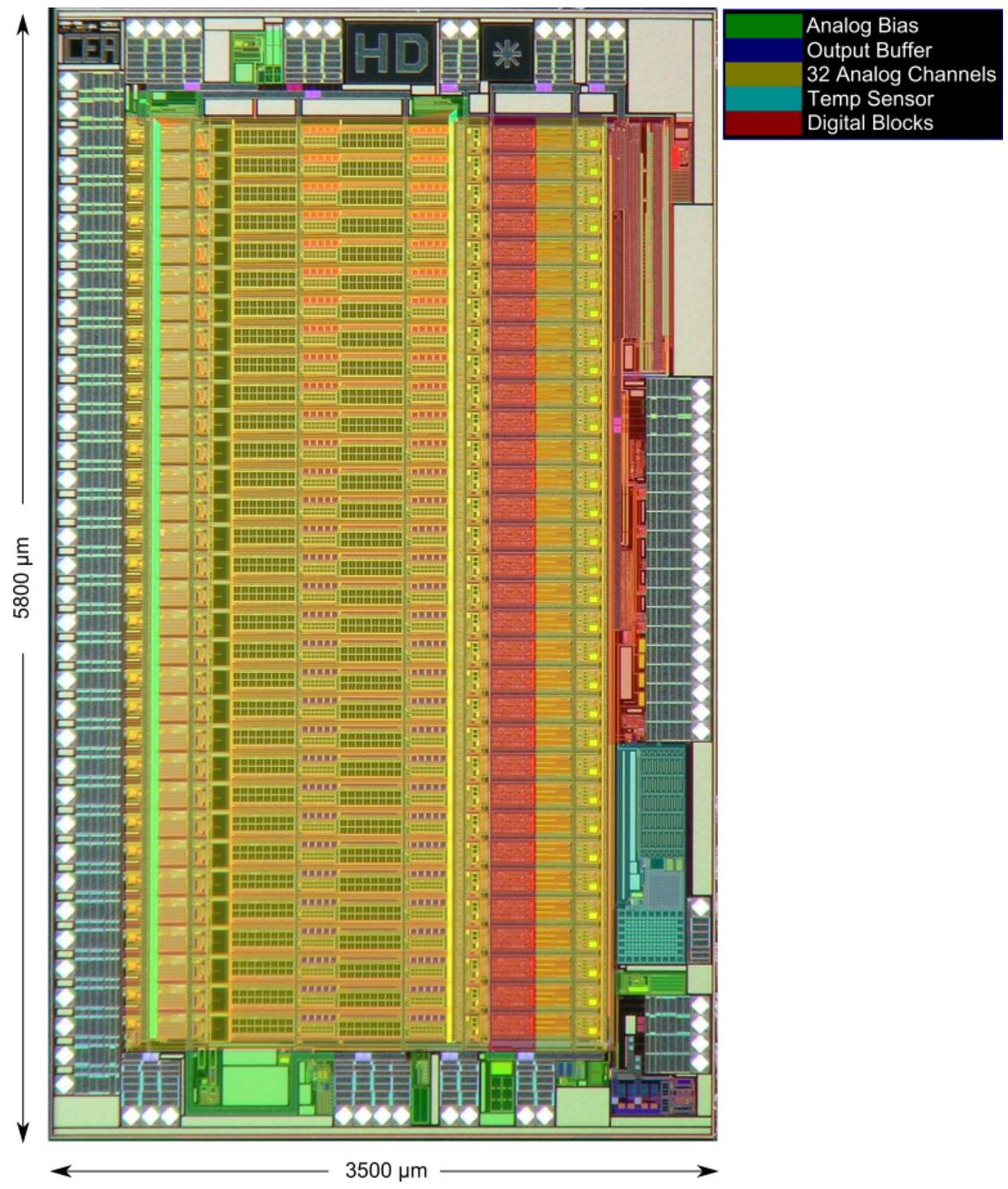

Fig. 2. Layout of the IDeF-X HD ASIC designed in CMOS AMS 0.35µm with SEL hardened digital library.

## 2.2 Channel design

### 2.2.1 Charge Sensitive Amplifier

The first stage of the analog channel includes a CSA biased at a total current of 50 µA, optimized for typical detector capacitance ranging from 2 to 5 pF with a continuous reset system performed by a PMOS transistor. The CSA has been optimized to be DC coupled to very low leakage current detector. However, it supports a leakage current up to 10 nA.

Compared to our previous designs [8, 11], we reduced the feedback capacitance from 200 fF down to 50 fF. In addition, we increased the input dynamic range up to 40 fC instead of 10 fC. Thus, the voltage dynamic range at the output of the charge amplifier has been increased by a factor of 16 and can reach an amplitude of 800 mV. By using a continuous reset operated by a MOS transistor in parallel with the feedback capacitance, one can considerably degrade the energy resolution by injecting a non-stationary noise (NSN). This noise is generated by the increase of the drain current in the reset transistor during the reset phase [18]. To delay this noise source, we integrated a low pass filter that we called Non stationary Noise Suppressor (NSNS) between the output of the charge amplifier and the source of the reset transistor.

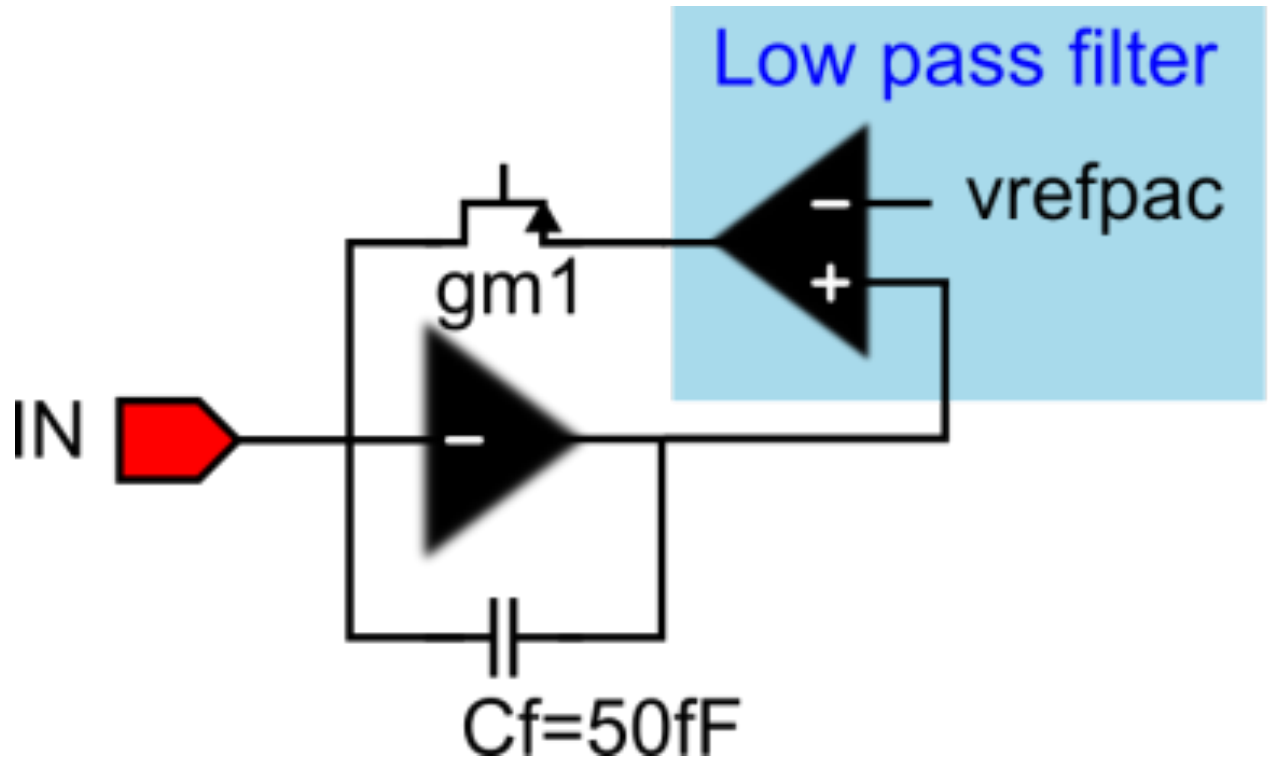

Fig. 3. Schematic of the charge sensitive amplifier: The reset loop integrates a low pass filter between the output of the amplifier and the source of the reset transistor to reduce the non-stationary noise.

Fig. 3 shows the schematic of the charge preamplifier and the low pass filter in the continuous reset loop. Thanks to a very high equivalent resistance designed with an OTA biased at very low current (few pA), the bandwidth of the low pass filter can be set to 100 mHz such that the drain current of the reset transistor will stay nearly unchanged during the pulse. In addition, the DC output of the charge preamplifier is stabilized at *vrefpac* and no longer depends on the leakage current of the detector.

The first plot of Fig. 4 shows the response of one channel at the output of the shaper (*Outfilter*) to a 27.5 fC injected charge with a ~20 pA leakage current and a peak time of 10.7 µs. This response has been recorded one thousand times by using an external 14 bit, 100 MSPS ADC. The delay between two consecutive acquisitions was set to 200 ms to prevent any correlation between consecutive acquisitions.

On the lower part of Fig. 4 we plotted the 1000 normalized signals to illustrate NSN effect, corresponding to a noise increase during the signal formation: for each time abscissa, we superimposed 1000 independent pulse height samples normalized to their mean. The measurements have been performed with (black dots) and without (grey dots) NSNS low-pass filter in the loop.

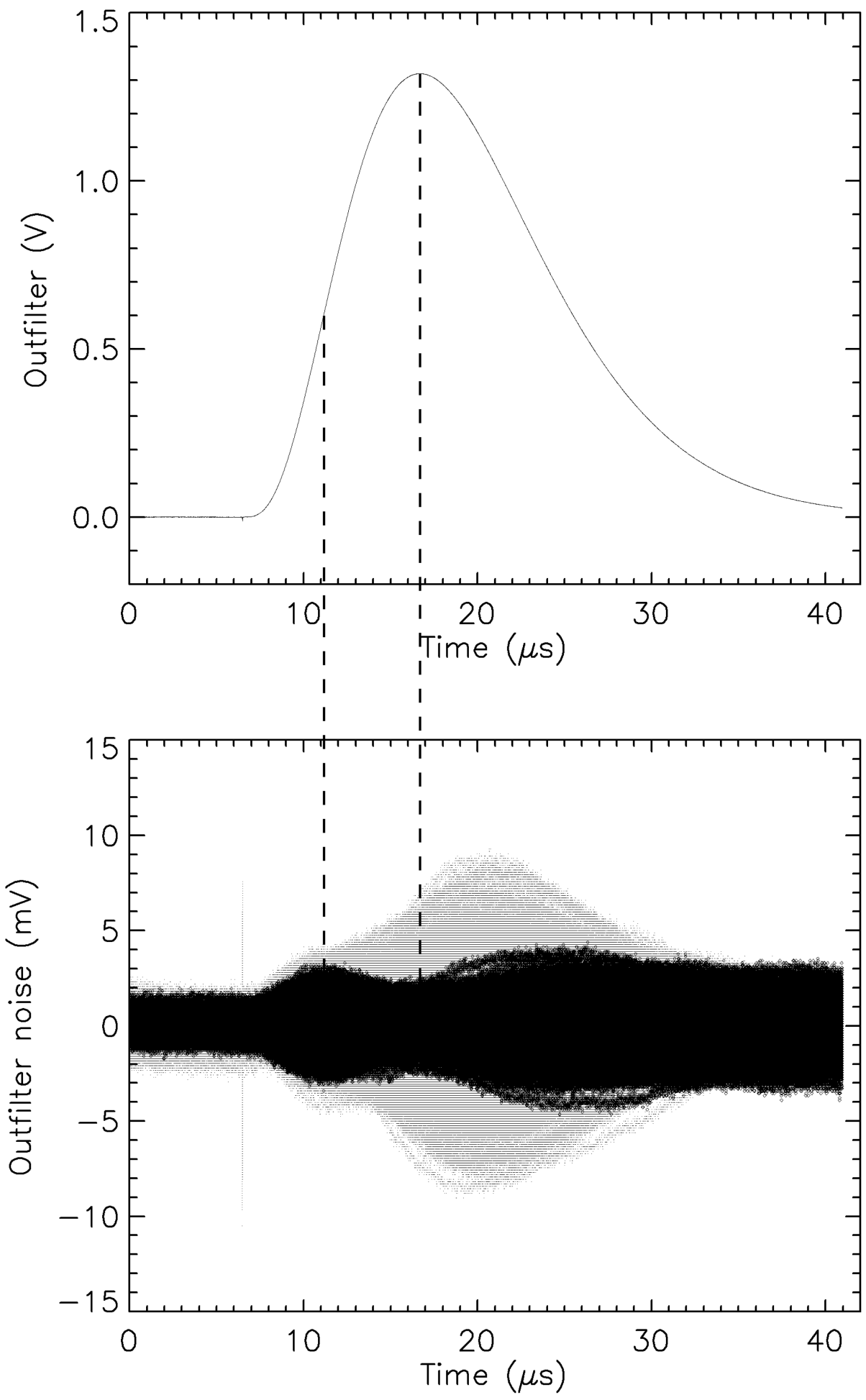

Fig. 4. Response to a 27.5 fC charge. a) Shape of the signal at the output of the channel. b) Noise during signal with and without NSNS.

One can see that in both cases, the noise increases during signal shaping. However, thanks to the low pass filtering, the noise at maximum is significantly reduced by the NSNS.

Careful interpretation of this plot is required as it could be subject to discussion. As a matter of fact, the precision of the measurement relies on the time accuracy of the acquisition trigger. In our setup, the signal that triggered each acquisition was the charge injection signal itself and was not the trigger of the ASIC. Thus, the relative delay between two acquisitions (jitter) could be as long as a sampling period, which corresponds to 10 ns when using a 100 MSPS ADC. Two consecutive acquisitions may be shifted by 10 ns. This shift has no influence on flat signals and does not affect the response on the baseline nor on the maximum of the shaped signal. Conversely, this effect is increased when the slope of the signal is high. In the case of Fig. 4, the slope is maximum around 11 µs and its value is 185 µV/ns with the 27.5 fC injected charge. Consequently, the time shift generates a maximum voltage shift of 1.85 mV. This is most probably the reason why we observe an increase of the noise when the slope of the signal is maximal. Possibly, this is an artifact. Nevertheless, the effect of the filter is clearly observable, efficiently saving noise when the peak detector will record the signal amplitude.

Making these measurements at different injection levels, one can plot (Fig. 5) the Equivalent Noise Charge of the channel at peak time versus the injected charge, with and without filter. Each point is derived from 1000 pulses. NSN is even more severe when the leakage current is high. We performed the comparative measurements in a worst-case setting, 20 pA leakage current, considered to be a fairly high value in our applications.

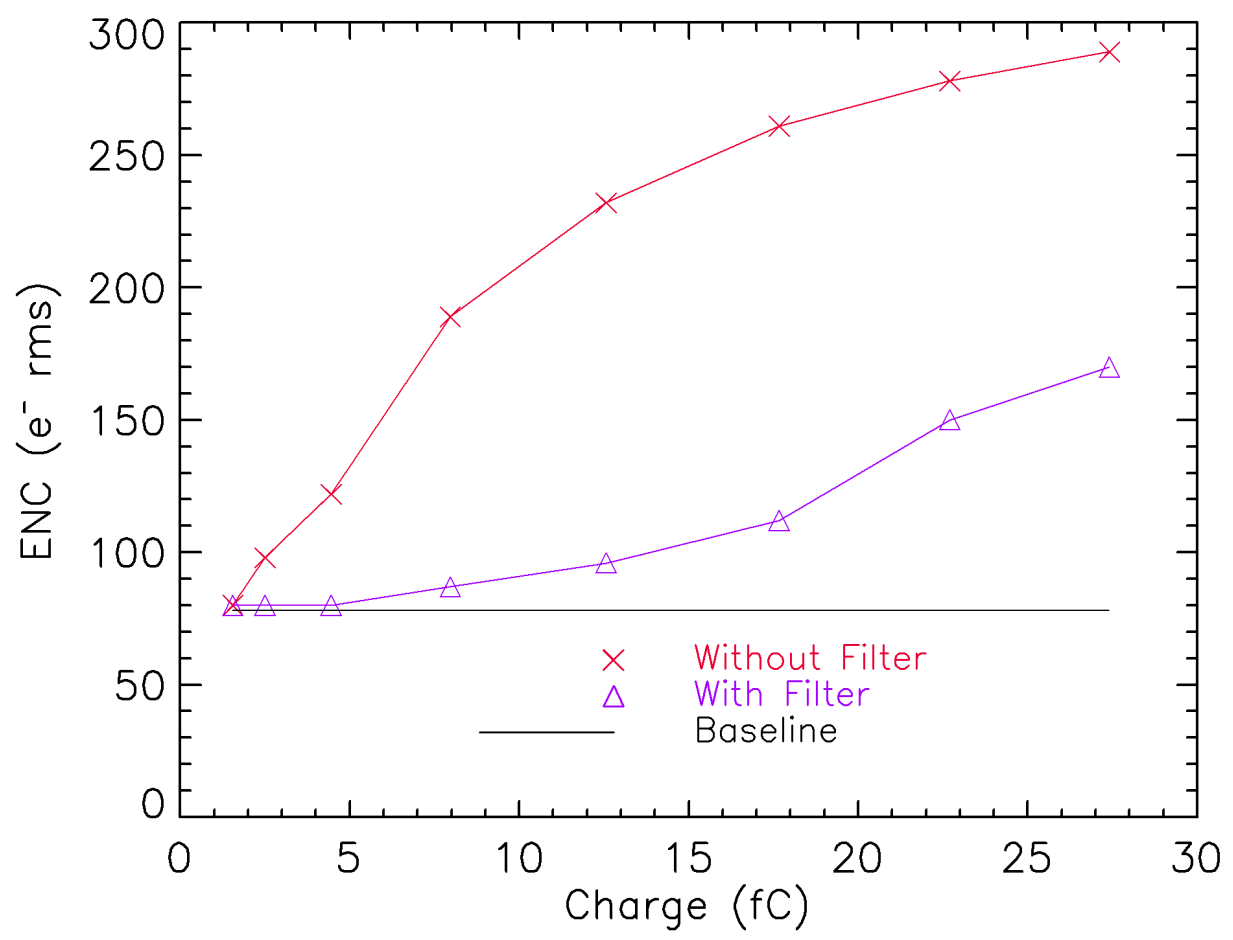

Fig. 5. Equivalent Noise Charge: influence of the injected charge with and without filter with a 20 pA leakage current.

Using a 2 pA leakage current, close to our applications expectations, we produced Fig. 6. We show the energy resolution derived from the baseline ENC and derived from the ENC measured on the pulse height at peak time using the NSNS function. The same resolutions are combined with the statistical fluctuation of pair creation in the CdTe assuming a Fano factor of 0.14 to extract the resolutions that we expect when connecting the ASIC to a CdTe detector. We conclude that the non-stationary noise is no longer affecting the spectral response up to 500 keV.

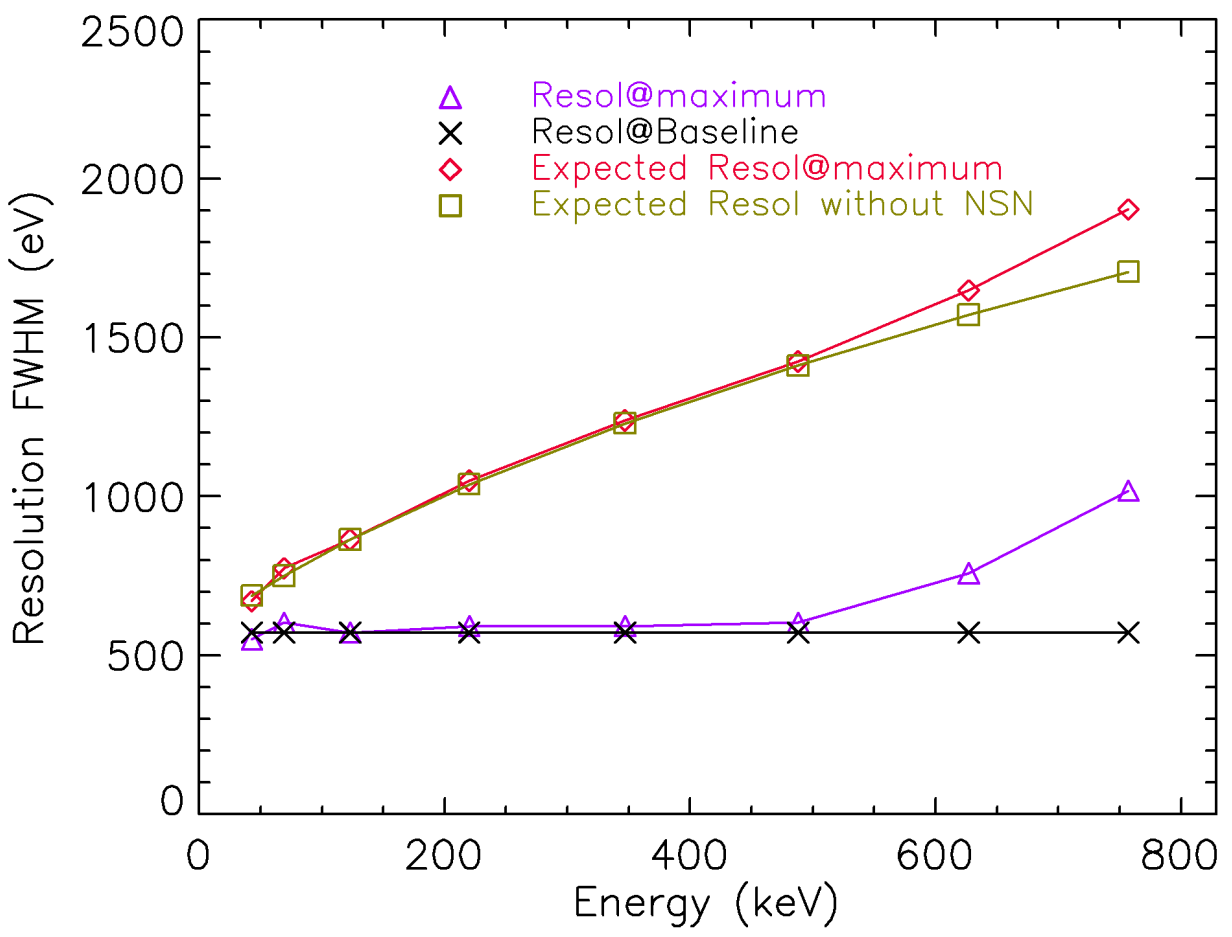

Fig. 6. Energy resolution: influence of the energy with NSNS with a 2 pA leakage current with and without detector noise.

### 2.2.2 *Pole-zero cancellation*

In previous design, we used a pole-zero cancelation architecture with continuous reset [17], that is now classical. In our new design we use the same kind of pole-zero cancellation technique but we introduce a simple resistor that adds a pole in the transfer function. Doing so, it is possible to increase the order n of the CR-RC$^n$ filter without any additional active device and consequently saving power.

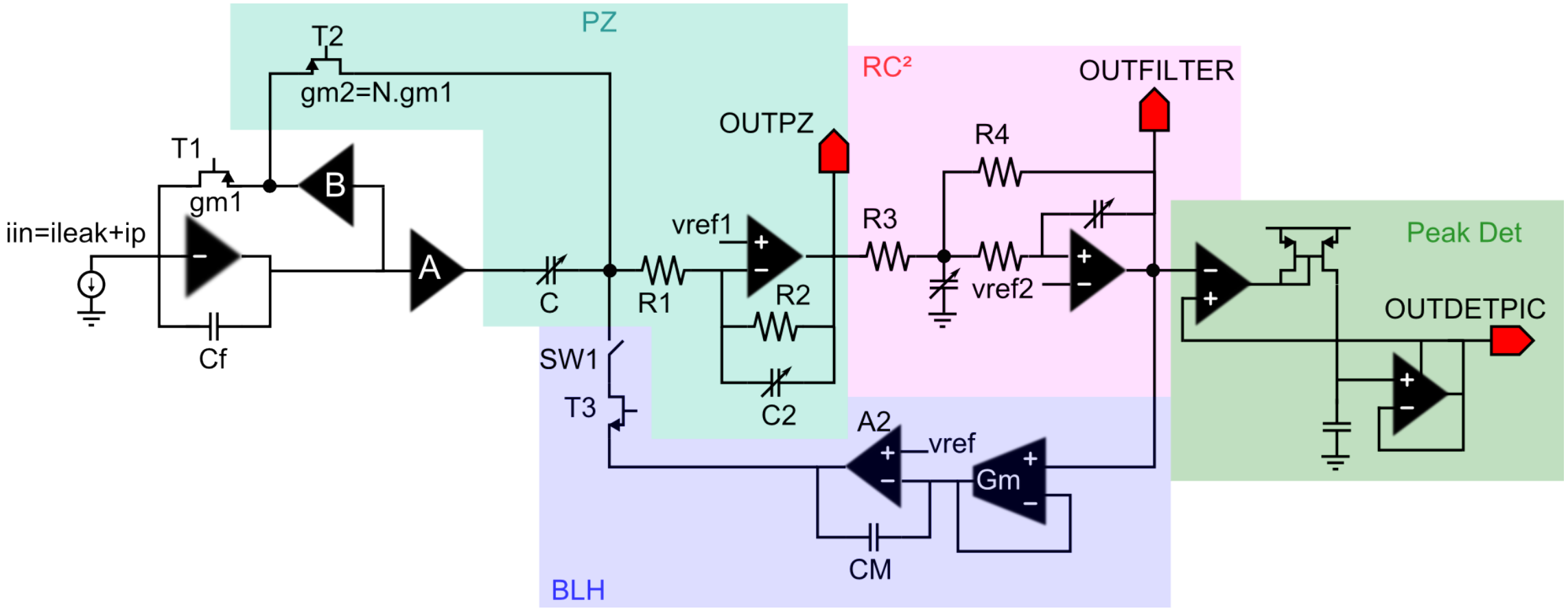

Fig. 7. Schematic of the energy path of the channel. The discrimination path is not represented.

The schematic of the channel is depicted Fig. 7.

- A is the adjustable gain of the amplifier.
- B gain of the NSNS amplifier.
- gm1 is the transconductance of the reset transistor T1.
- N gain factor of the pole-zero cancellation stage.

The current pulse *ip* from the detector is converted into voltage by the CSA and amplified by the adjustable gain amplifier A. The variable dominant pole $Bg_{m1}/C_f$ is cancelled by C and T2 that is equivalent to N×T1 PMOS transistors.
The frequency response of the output of the pole-zero cancellation stage to a pulse *ip* is:

$$OUTPZ = -\frac{R_2 N\left(1 + j\dfrac{AC}{BNg_{m1}}\omega\right)}{\left(1 + j\dfrac{C_f}{Bg_{m1}}\omega\right)(1 + jR_1C\omega)(1 + jR_2C_2\omega)} ip$$

In previous architectures, $R_1 = 0$ and the pole cancellation is done by setting $C = \frac{NC_f}{A}$. The response becomes:

$$OUTPZ = -\frac{R_2 N}{(1 + jR_2C_2\omega)} ip$$

With our current design, by setting $C = \frac{NC_f}{A}$ and introducing an additional resistance $R_1 = \frac{R_2 C_2}{C}$ , the response becomes:

$$OUTPZ = -\frac{R_2 N}{(1 + jR_2C_2\omega)^2} ip$$

Thus, the order of the filter at the output can be doubled without any additional active device and any additional power consumption, just adding the $R_1$ resistor. Since $R_1$ and $R_2$ are fixed, the two variable parameters of the pole zero cancellation stage are the gain A and the shaping time ts. All the other parameters can be expressed with the following variables:

$$\begin{cases} C_2(t_s) = \dfrac{t_s}{R_2} \\ C(t_s) = \dfrac{t_s}{R_1} \\ N(t_s, A) = \dfrac{At_s}{R_1 C_f} \end{cases}$$

Modeling the input charge by a Dirac pulse with a $Q_{in}$ value, the transient response at the output of the pole zero stage can be easily calculated:

$$OUTPZ(t) = -\frac{R_2 A}{R_1 C_f} Q_{in} \frac{t}{t_s} e^{-\frac{t}{t_s}}$$

The signal peak is reached at *t=ts* and its value is:

$$OUTPZ_{peak} = -\frac{R_2 A}{R_1 C_f} Q_{in} e^{-1}$$

The gain remains independent of the shaping time and only depends on the adjustable amplifier gain *A*. Fig. 8 shows the normalized response $OUTPZ/OUTPZ_{peak}$ as a function of *t/ts* for different values of *ts*. The different curves should coincide and reach their maximum value at *t/ts* =1. This is true for *ts* >1µs. For lower values, the gain decreases, and the time at maximum is higher than its theoretical value. While the decrease of the gain is not yet fully understood, the higher time at maximum for very short *ts* values can be explained. In fact, in the calculation, we considered the response to a perfect Dirac while the real signal is coming from the CSA. Due to the rise time at the output of the CSA (approximately 50 ns), the signal at the output of the PZC stage is delayed and this is the reason why we observe such a higher time at maximum, especially at short *ts*.

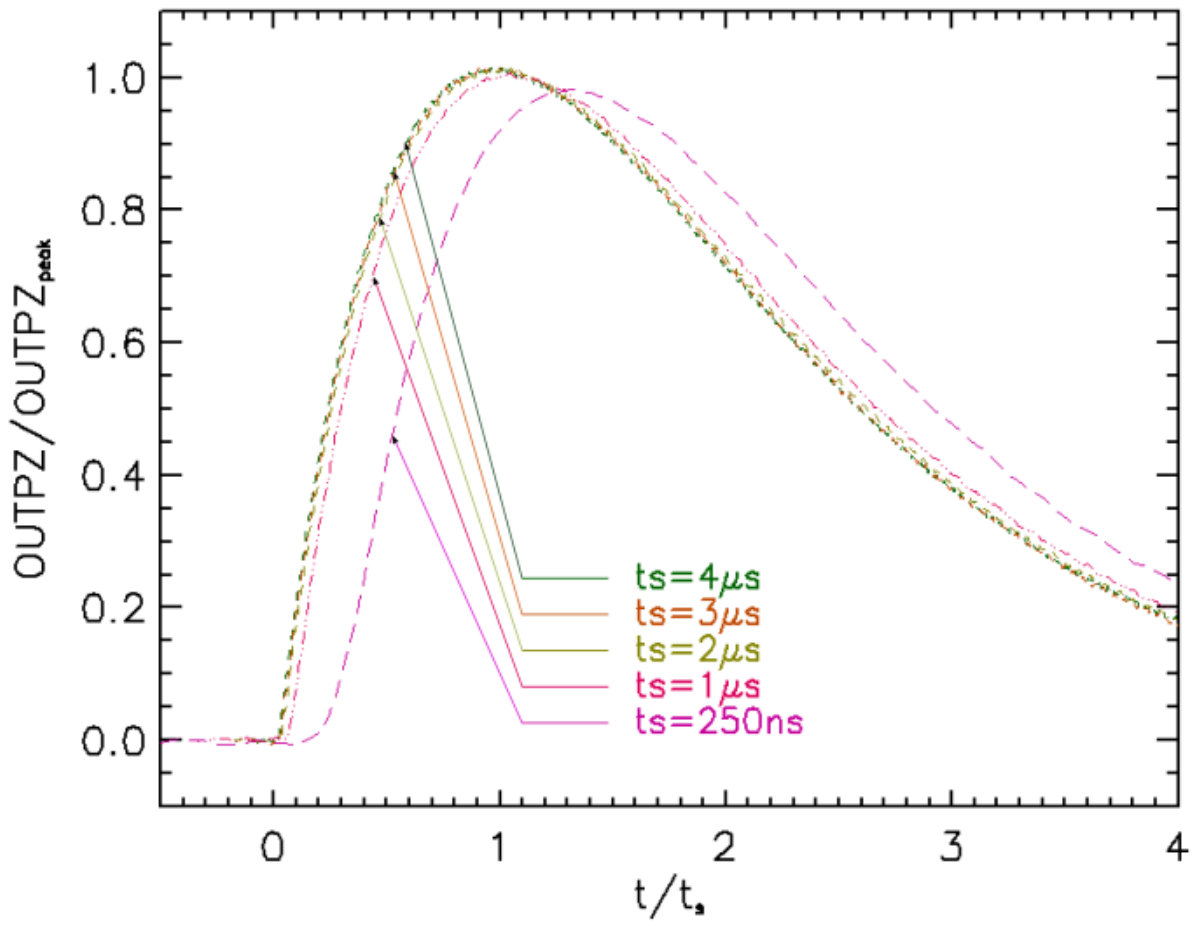

Fig. 8. Response of the pole zero cancellation stage as a function of t/ts.

### 2.2.3 *Base line stabilization (BLH)*

The baseline stabilization is performed by a low pass filter inserted in a feedback loop between the output of the RC² filter (OUTFILTER) and the input of the PZ stage (Fig. 7). This filter is a slow transconductor that compensates for variations of the amplified leakage current coming from the transistor T2. It ensures that the current remains constant in the following resistors (R1-4). The transconductor is designed with a very low bias current (1 nA) OTA (Gm) that charges the Miller capacitance A2*CM of a voltage amplifier. The output of the amplifier is connected to the source of a NMOS transistor (T3) and thus commands the amount of current that has to be released to T2. The DC output of the channel (OUTFILTER) can be tuned by the voltage reference (VREF) of the amplifier which is delivered by a common 3-bits DAC. In addition, the loop can be opened thanks to the SW1 switch. When the switch is open, the whole channel amplifies the input leakage current and can be used as a pico-Amp meter.

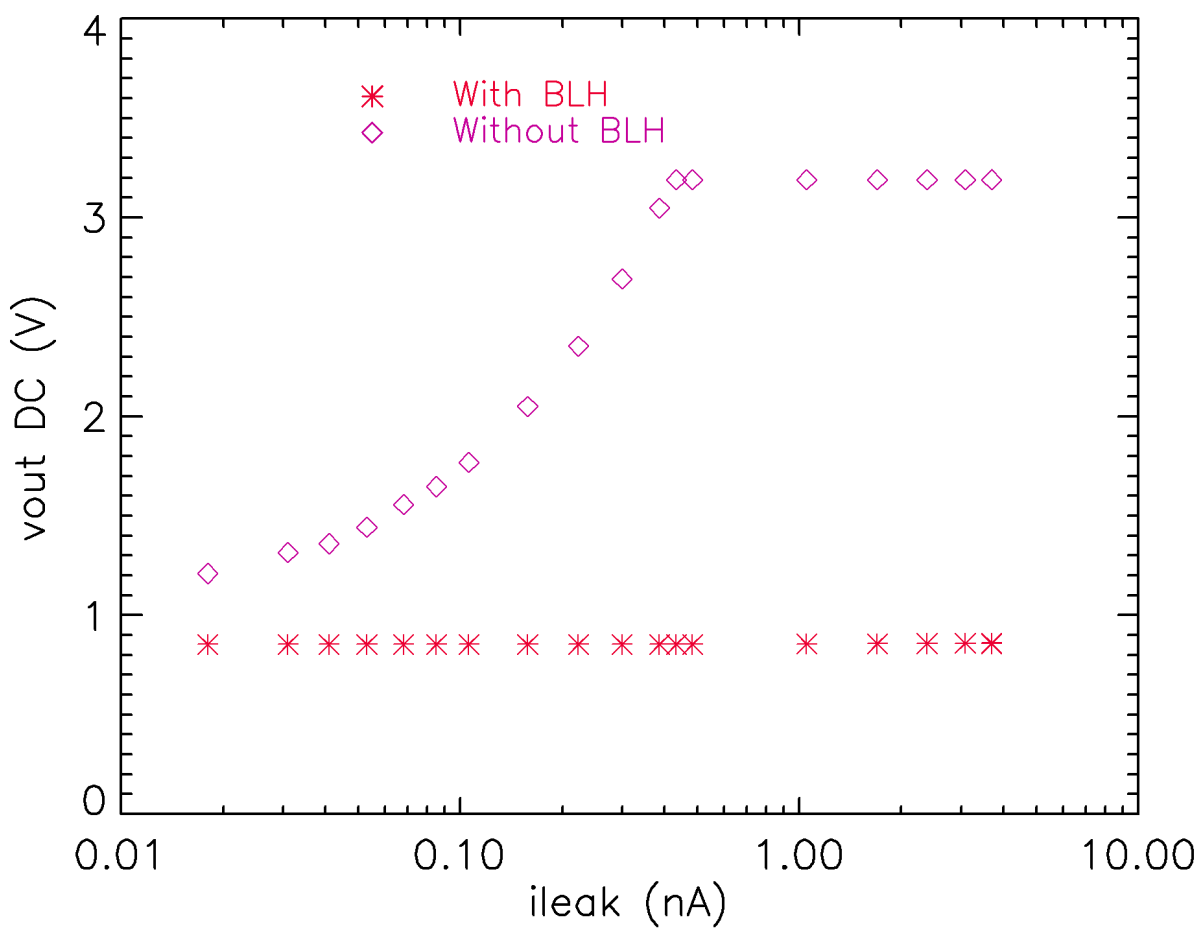

Fig. 9. Influence of the input leakage current on the DC output of the channel with and without baseline holder.

Fig. 9 shows the measured DC output of the channel as a function of the input leakage current with and without BLH. When the BLH is on, one cannot see any influence of the current on the DC output up to more than 3 nA. Consequently, a detector offset calibration is independent of the detector current which is subject to change with the operational temperature variations for instance. For all the following measurements that are presented in the paper, the BLH is always activated.

# 3 General Electrical Measurements results

## *3.1 Functionality*

The functionality of processing channel is presented in Fig. 10. Shaped signal due to input charge injection was recorded for three values of peak time at 2.72, 6.73 and 10.73 µs. The signal was observed directly at the output of the filter (OUTFILTER) with gain set internally to 200 mV/fC. Fig. 10 shows that gain remains constant whatever the selected peak time. Moreover, the PZ-cancelation stage operates properly since no undershoot or tail is observed at all.

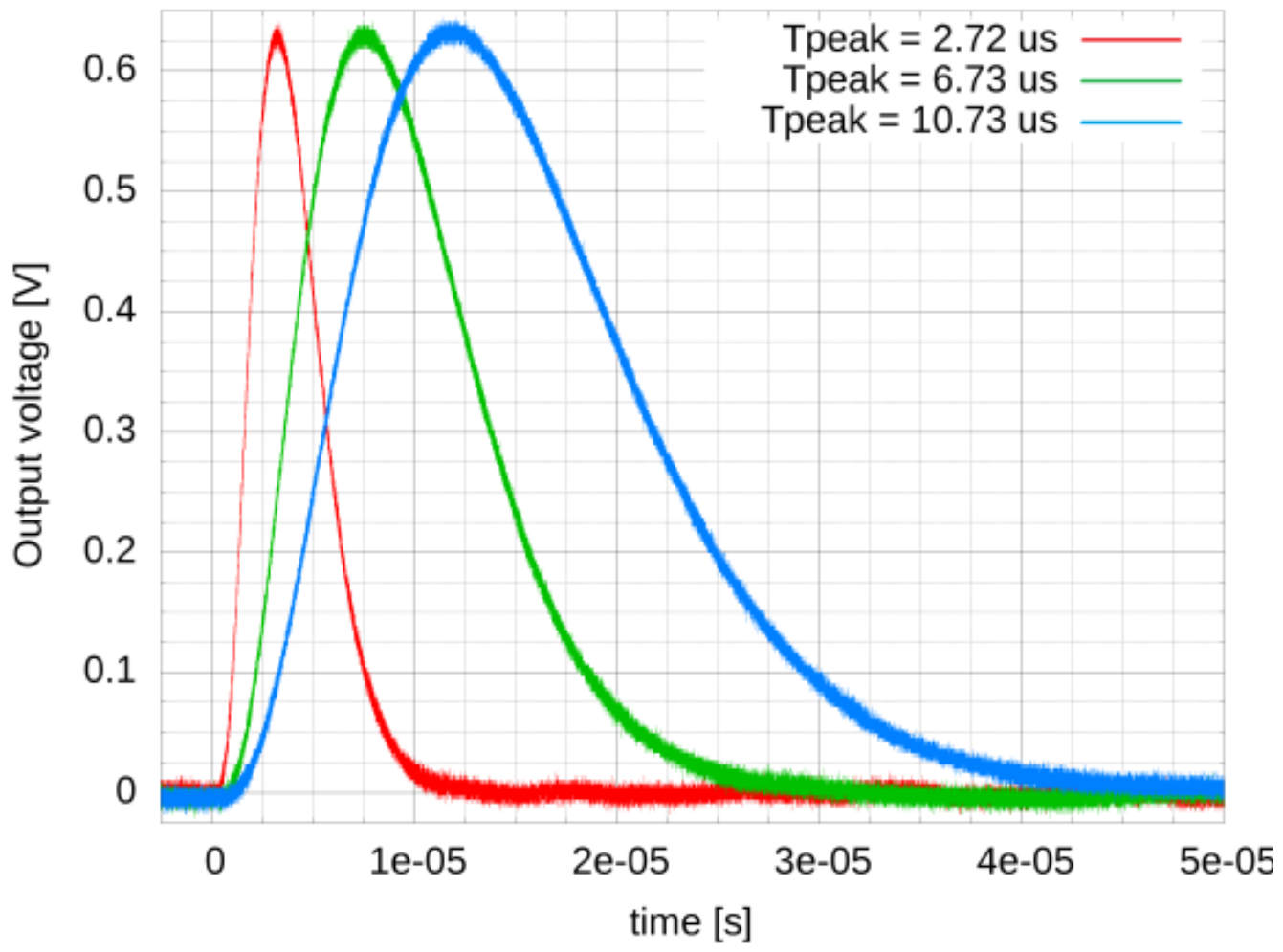

Fig. 10. Shaped signal at the output of the shaper when injecting a 3 fC input charge for three values of peak time: 2.72 µs, 6.73 µs and 10.73 µs with gain 200 mV/fC.

### *3.2 Gain and Linearity*

Fig. 11 shows the transfer function at the output of the RC$^2$ filter of the channel #31 for the four available gain values 200, 150, 100 and 50 mV/fC at a peak time of 10.73 µs. The dynamic range enables observing energy signals up to 0.3 MeV, 0.37 MeV, 0.6 MeV and 0.99 MeV with a CdTe detector. Furthermore, integral nonlinearity (INL) was measured and illustrated in Fig. 12 for 50 mV/fC and 200 mV/fC gains. When the input charge remains lower than 10 fC, the INL remains in a ±0.6 % window, while it will keep within ±1.2 % for higher dynamic ranges.

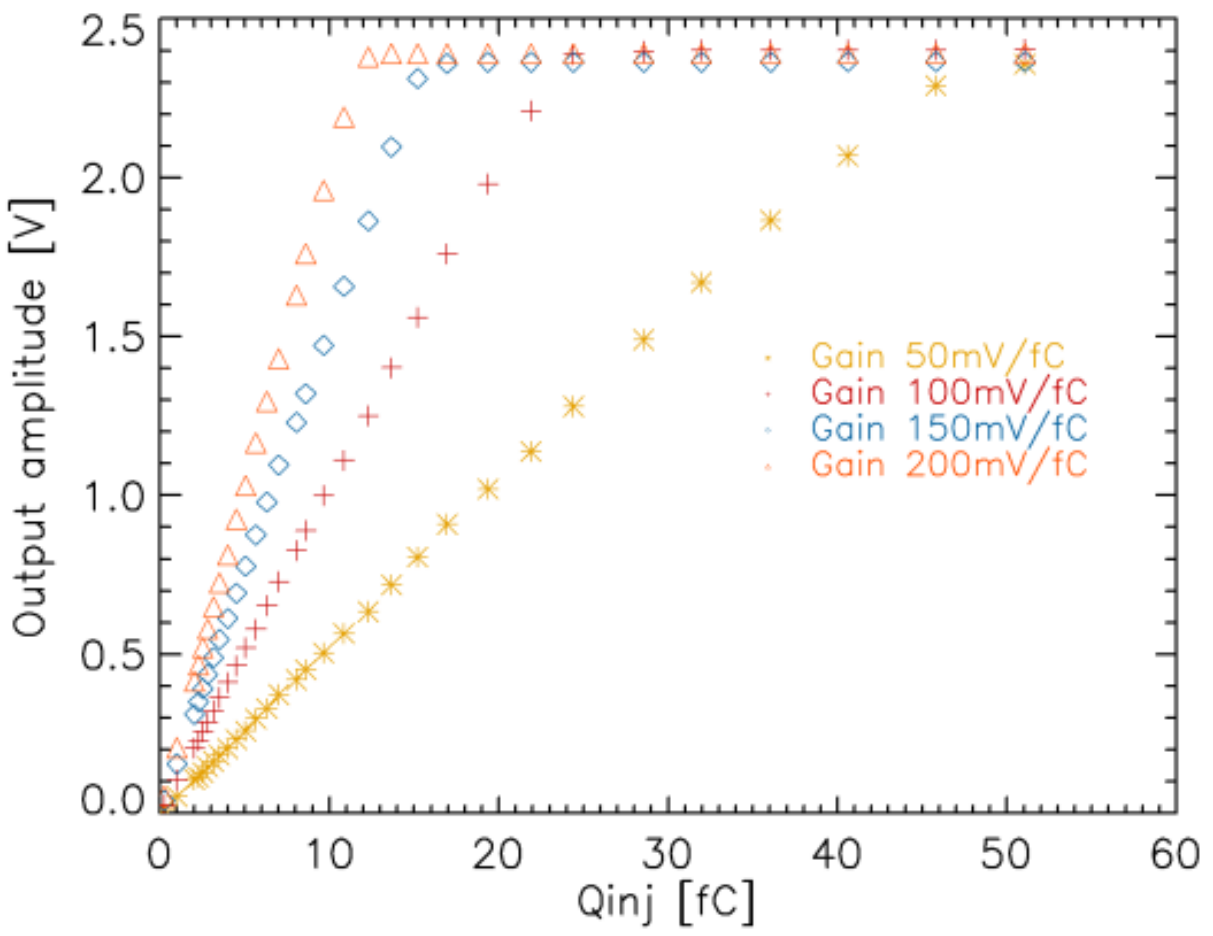

Fig. 11. Transfer function at the output of the RC$^2$ filter for 4 values of gain. With the low gain configuration, the dynamic range goes above 36 fC (0.99 MeV for CdTe).

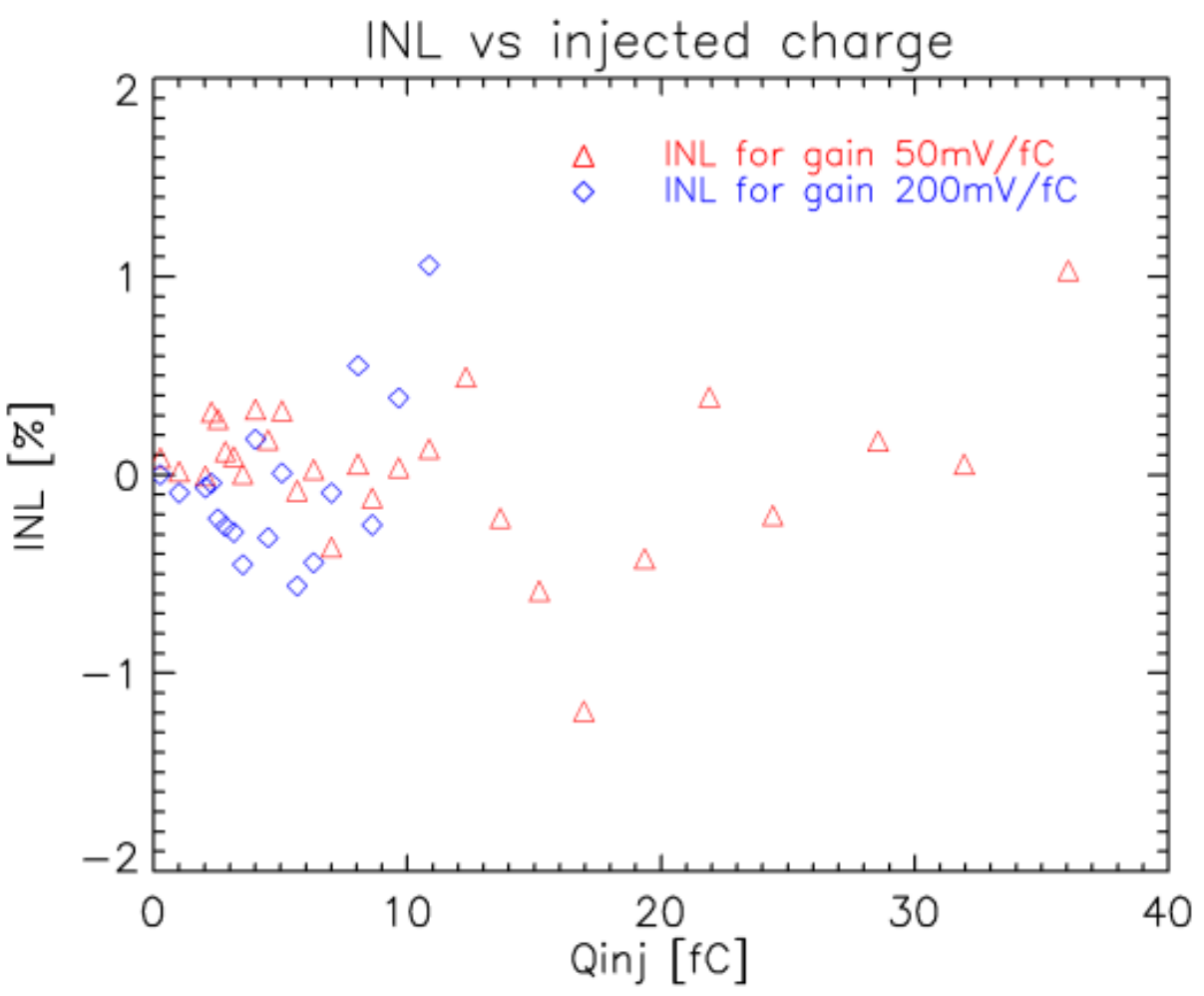

Fig. 12. Integral nonlinearity for gain 200 mV/fC and 50 mV/fC with dynamic range respectively 11 fC and 36 fC.

### 3.3 *Equivalent Noise Charge*

Since the IDeF-X HD chip has been designed to be connected to high spectral resolution detectors (~1 keV FWHM at 60 keV), a key characterization of the circuit consists in measuring the Equivalent Noise Charge (ENC) of the channel. In order to gather full characteristics of the ASIC and to build the noise model of a channel, we measured the noise response for a range of input capacitances and for a range of input leakage currents.

Fig. 13 presents the ENC measurement results for different values of capacitances ranging from 0 to 34 pF, soldered at the input of channel 31. The sensitivity to leakage current at the highest peak time is depicted in Fig. 14.

When no detector is connected to the chip and no leakage current is programmed (ileak ~1pA), the noise floor is found to be 33 $e^-$ rms. This value is similar to the ones obtained with previous versions of IDeF-X chips but is achieved with total power consumption reduced by a factor of 4.

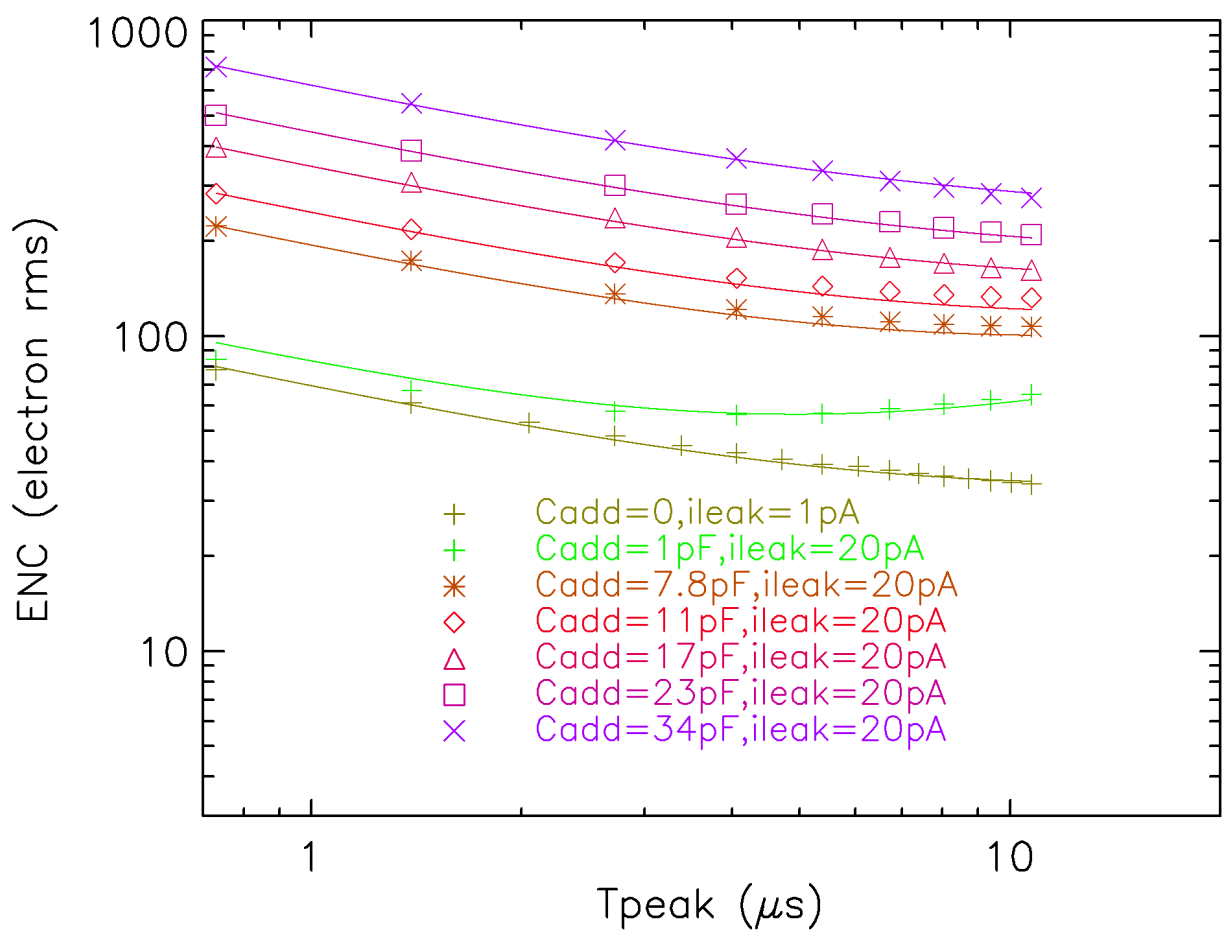

Fig. 13. ENC as a function of peak time: measurement and model fit. Gain 200 mV/fC, leakage current 20 pA.

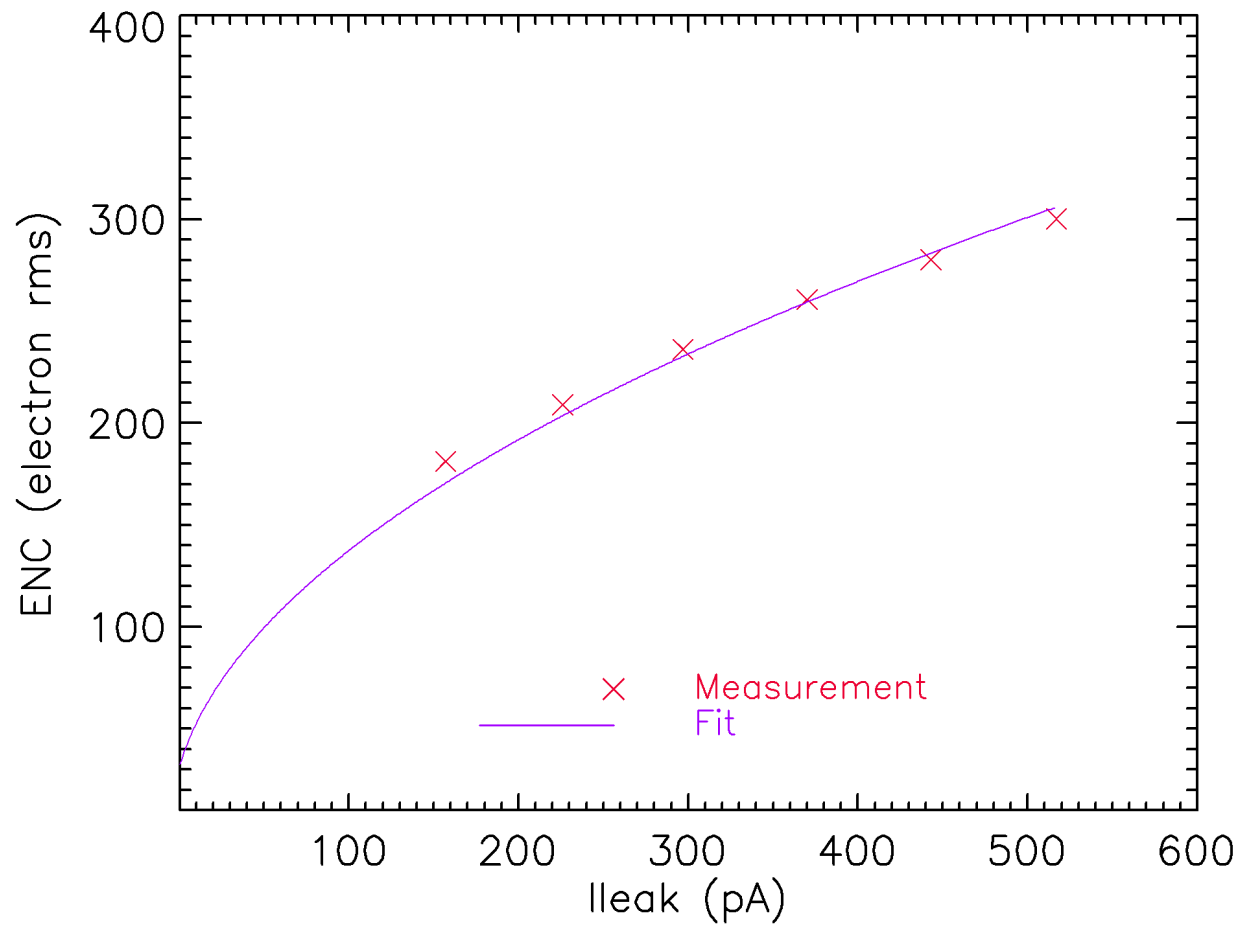

Fig. 14. ENC as a function of $I_{leak}$ without additional input capacitance: measurement and fit. Gain 200mV/fC, Tpeak=10.73µs.

The theoretical expression of the ENC at the output of the shaper without non stationary noise is [19]:

$$ENC^2 = C_{tot}^2\left(\frac{\alpha_d}{t_{peak}} + \alpha_{1/f}\right) + \alpha_{//} \cdot i_{leak} \cdot t_{peak} \quad (2)$$

Where:

- $C_{tot} = C_0 + C_{add}$,
- $C_0$ is the total capacitance at the input of the CSA including the stray capacitance of the ASIC board,
- $C_{add}$ is the capacitance of the additional test capacitor soldered at the input,
- $\alpha_d$ is a parameter that depends on the filter order as well as the transconductance of the input transistor of the CSA,
- $\alpha_{1/f}$ is a parameter that depends on the filter order as well as technological parameters and the area of the input transistor,
- $\alpha_{//}$ is a parameter that depends on the filter order,
- $i_{leak}$ is the leakage current injected at the input of the CSA.

It is possible to extract $\alpha_d$ , $\alpha_{1/f}$ and $C_0$ parameters fitting the ENC response plotted on Fig. 13. Doing so, we derive $\alpha_{//}$ using these parameters to fit the noise sensitivity to leakage current as shown on Fig. 14. The complete set of noise parameters is given in Table I.

TABLE I

IDEF-X HD NOISE PARAMETERS

| PARAMETER | MEASURED VALUE |
|---|---|
| $\sqrt{\alpha_d}$ | 488 $e^-.ns^{1/2}.pF^{-1}$ |
| $\sqrt{\alpha_{1/f}}$ | 5.67 $e^-$/pF |
| $\alpha_{//}$ | 16.7 $e^-.ns^{-1}.nA^{-1}$ |
| $C_0$ | 3.9 pF (ASIC) + 0.3 pF (stray) |

Extracted parameters enable to calculate ENC for any capacitance and leakage current. In our Caliste application, we expect the total stray capacitance at each individual input of IDeF-X HD, including pixel, interconnections and surrounding materials stray capacitances, to be 2 pF typical, while the pixel dark current of cooled detector will be 20 pA maximum. These values are considered to be our worst-case scenario. It led to a minimum ENC of 68 $e^-$ rms at 5.4 µs peak time. Thus, the objective of 1 keV at 60 keV is easily achievable.

# 4 Spectroscopy Measurements

## *4.1 Connection to CdTe mono pixel*

In order to perform preliminary spectral evaluation and demonstration, the chip was connected to a single Schottky CdTe detector equipped with Al blocking contact at the anode (Acrorad, Japan) and Pt quasi-ohmic contact at the cathode. The anode is patterned with a 2 × 2 mm² pixel surrounded by a 1 mm guard ring separated by a 50-micron gap. The detector is 4.1 × 4.1 × 2 mm³. During measurements, the detector was illuminated at the cathode side. The Al pixel was connected to one of IDeF-X HD channels (channel #31). The reverse voltage applied to the diode was set at –1020 V. We cooled down the crystal to approximately at –10 °C. The guard ring voltage was set to the input voltage of the chip to avoid surface currents between the pixel and the guard ring. The detector is mounted onto a thin ceramic board and connected to the chip through a TO package to facilitate operations. Thus, the overall assembly represents approximately a 5 to 10 pF additional stray capacitance at the input when the detector is fully depleted. On the other hand, the detector dark current at –1000 V reverse voltage has been measured at 25 °C and is ~170 pA. From this measurement, we extrapolate the current to be ranging from 4 to 7 pA at –10 °C.

First measurement was done with a $^{241}$Am source. Spectrum of Fig. 15 was obtained with a 200 mV/fC gain configuration setting a 10.06 µs peak time. The spectral resolution is found to be 1.01 keV FWHM and 1.11 keV FWHM at 13.94 keV and 59.54 keV respectively. Assuming a Fano factor of 0.14, we derive the electronic noise to be ~95 $e^-$ rms. The result is consistent with the one obtained from ENC measurements for an 8 pF stray capacitance and 10 pA dark current.

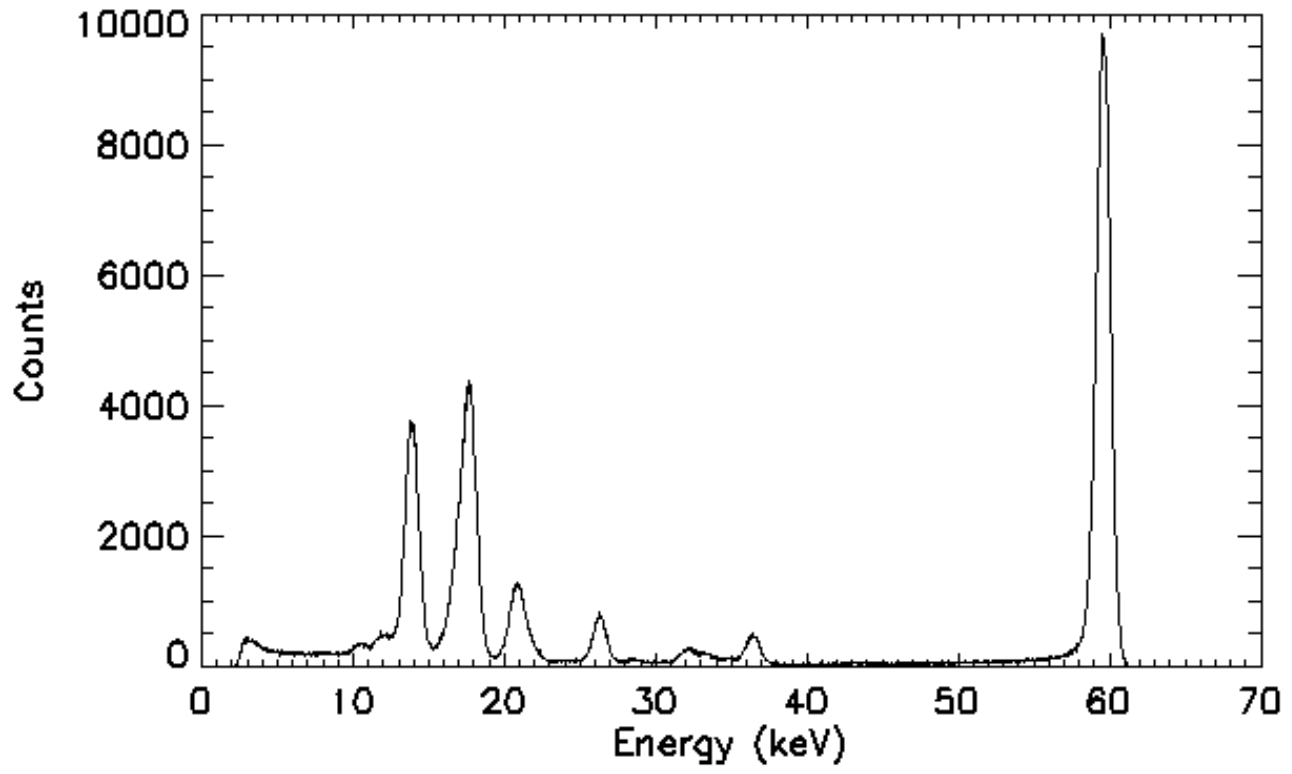

Fig. 15. Spectrum of the $^{241}$Am source obtained with 2×2×2 mm$^3$ CdTe single pixel at –10 ºC coupled to IDeF-X HD with a gain set to 200 mV/fC. The energy resolution is is found to be 1.11 keV at 59.54 keV and 1.01 keV at 13.94 keV.

A second spectroscopy measurement experience was done with a gain set to 50 mV/fC, still using a 10.06 µs peak time. Detector and configuration remained the same as before. We measured performance over high-energy range with $^{137}$Cs source (Fig. 16). The energy resolution was found to be 1.2 keV FWHM at 32.2 keV (Ba $K_\alpha$ fluorescence line doublet), 1.1 keV FWHM at 72.8 keV (Pb $K_\alpha$ fluorescence line) and 4.3 keV FWHM at 661.7 keV ($^{137}$Cs line). Note that we achieved a low threshold level setting down to 4 keV in this configuration. From our measurements, we derive an electronic noise of ~100 e$^-$ rms. However, this value is fully consistent with the spectral response at 662 keV and further investigations are needed.

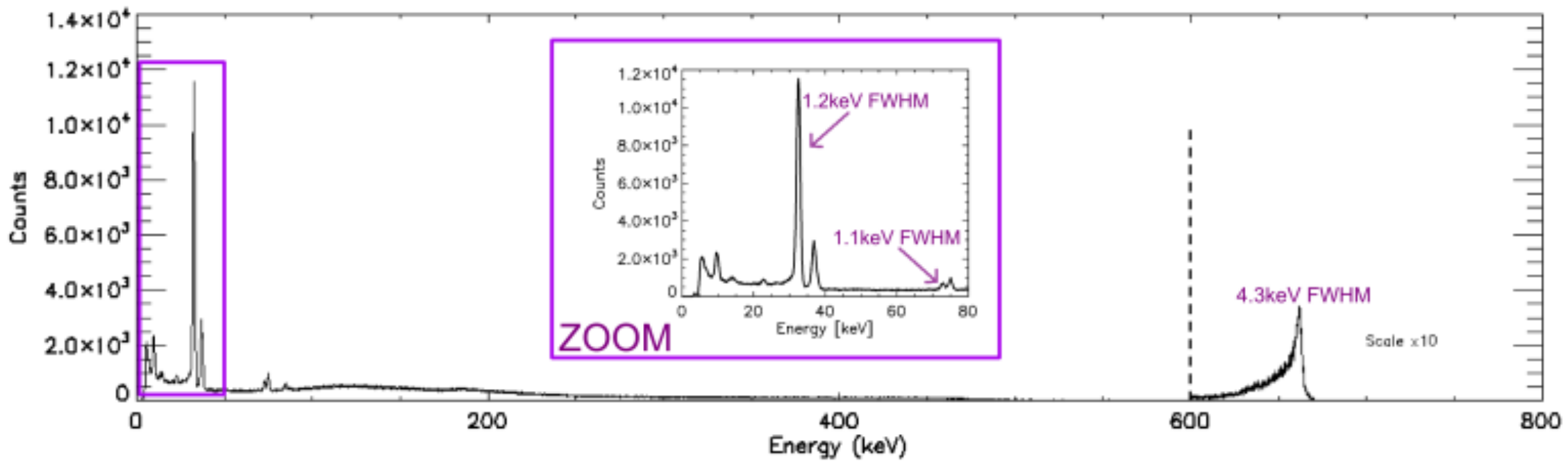

Fig. 16. Spectrum of the $^{137}$Cs source obtained with 2x2x2 mm$^3$ CdTe at -10 ºC coupled to IDeF-X with gain set to 50 mV/fC. The resolution is 4.3 keV at 662 keV, 1.1 keV at 72.8 keV and 1.2 keV at 32.2 keV (double peak).

### *4.2 IDeF-X HD in Caliste HD*

IDeF-X HD has been used to build Caliste HD, a new high spectral and high spatial resolution low power 256-pixels imaging spectrometer for hard X-rays. The Caliste concept has already been depicted in [13]. Fig. 17 shows two Caliste HD modules with different detector thickness.

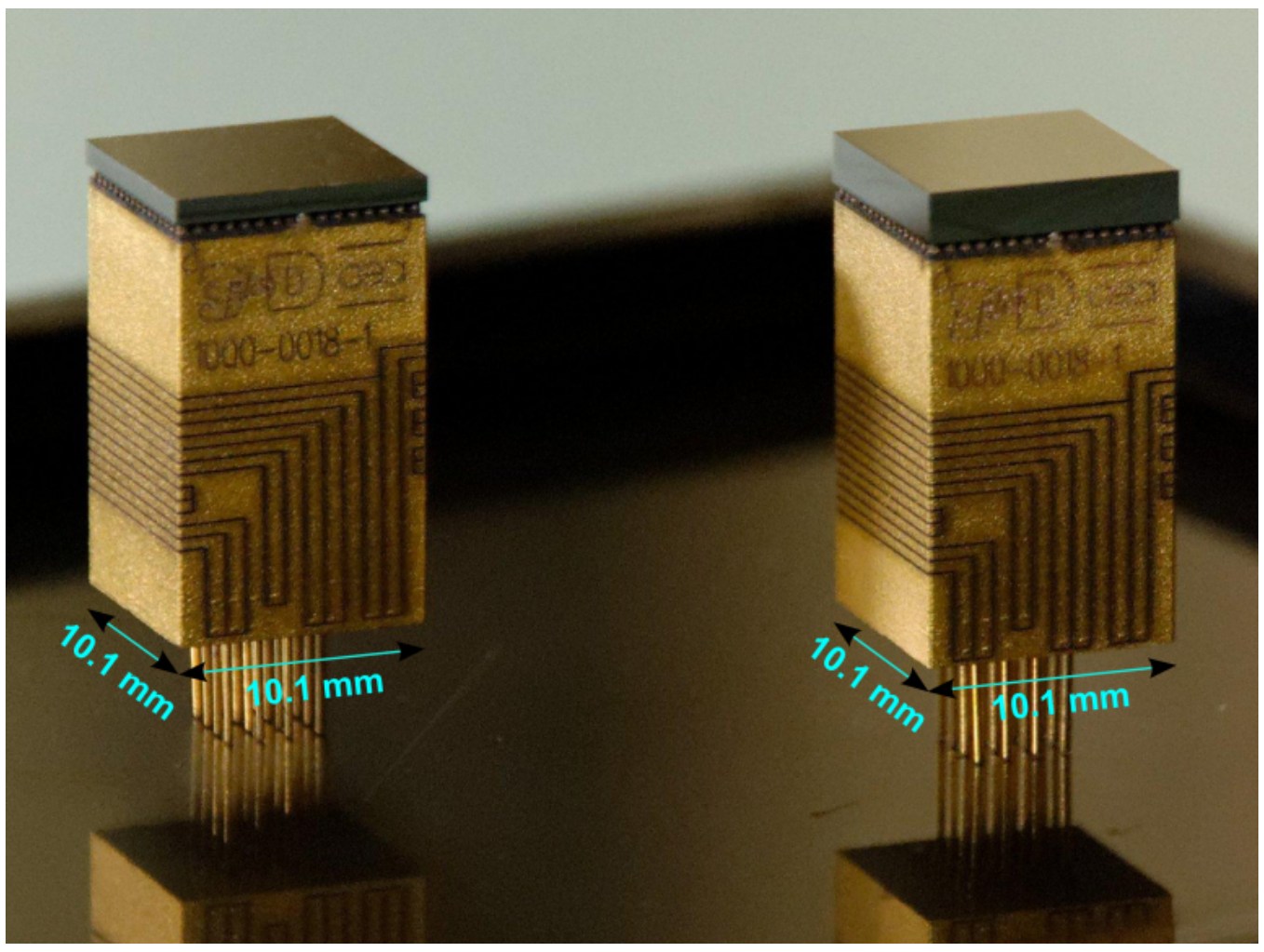

Fig. 17. Caliste HD imaging spectrometer units processed by 3D Plus with 1 mm-thick and 2 mm-thick Al/CdTe/Pt detectors from Acrorad.

In our Caliste HD device, eight IDeF-X HD ASICs are mounted on micro printed circuit boards called flexes. The eight flexes are then stacked together and molded into an epoxy resin to build the electrical body of the module. A 16 × 16 pixelated CdTe detector can then be glued on the top side of the electrical body.

Thanks to the new digital interface integrated in the chip, all digital signals can now be shared between ASICs in the module. Each ASIC has its own address that is coded by dedicated routing along the sides of the module. Each ASIC can then be read or programmed independently. Thus, the bottom interface of the module is made of a 1.27 mm pitch 4 × 4 pin grid array, instead of 7 × 7 in the previous Caliste modules. This compression has been performed to make the module easier to integrate in the complete camera MACSI [14].

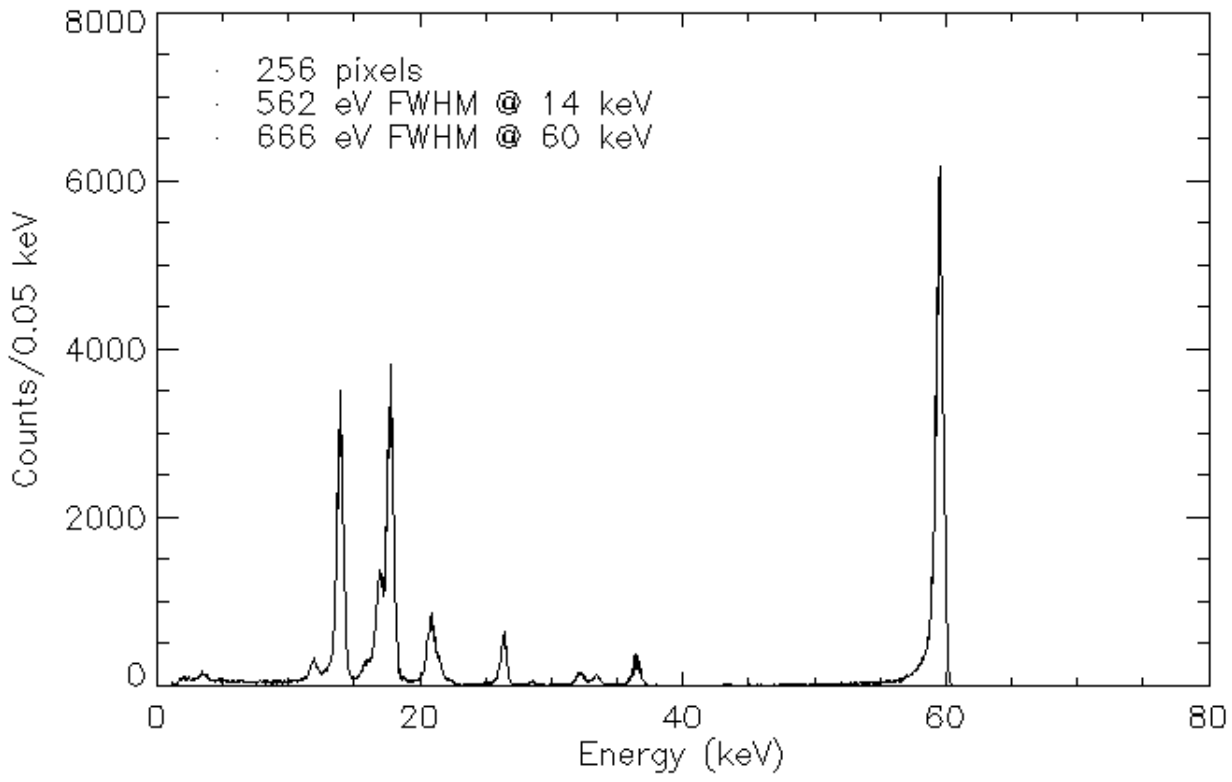

Fig. 18. Sum spectrum of the $^{241}$Am source on 256 pixels obtained with a 1 mm thick detector at -4 °C. The energy resolution is found to be 666 eV FWHM at 59.54 keV and 562 eV FWHM at 13.94 keV.

Fig. 18 shows the sum spectrum obtained with one Caliste HD module build with a 1 mm thick CdTe detector with a 625 µm pixel pitch biased at –400V. The sum spectrum was build using an individual pixel calibration table. Only single events have been recorded to create this figure. During the acquisition, the module was cooled down to –4 °C and exposed to an $^{241}$Am radioactive source.

The energy resolution is 562 eV FWHM at 13.94 keV and 666 eV FWHM at 59.54 keV. Pixels have low-level thresholds of ~ 1.2 keV. This very good spectral performance is due to the low noise frontend electronics associated with low input capacitance (< 1.5 pF) and very low leakage current of the small pixels of the Al Schottky CdTe detector (< 10 pA). Since our last paper on Caliste HD [13], we improved our acquisition boards and measurement environment to reduce all additional noise sources; we thereby achieved a significantly better energy resolution even at higher temperature.

# 5 Radiation effects

The space radiation environment can induce many degradations (dose effects), dysfunction (SEU, Single Event Upset) or even destruction (SEL, Single Event Latch-up) of the CMOS integrated circuits [20]. Since the IDeF-X HD ASIC is designed for space-borne application in astrophysics, we evaluated its sensitivity to radiation and to heavy ions as part of a space qualification process.

## *5.1 Dose effects*

We irradiated one IDeF-X HD circuit with a 589 GBq $^{60}$Co source during 7 months. In order to irradiate the chip at a very low dose rate quite similar to the one expected during a typical space mission, the ASIC was placed 45 cm from the source and was thus exposed to a 36 rad/h dose rate up to a 100 krad accumulated dose. Later, we placed the ASIC 13 cm from the source to increase the dose rate up to 200 rad/h. We performed a set of fourteen irradiation tests starting from 1 up to 200 krad of accumulated dose. During irradiation, the chip was biased with a 100 pA compensating current to ensure its functionality during the entire irradiation campaign and to avoid dysfunction due to an increase of the input protection reverse current [10]. During each test, we measured the gain, power consumption and ENC of the IDeF-X HD ASIC.

*1) Gain:* We measured the gain at all peaking times for each irradiation test. We injected a charge of 1 fC through the 50 fF injection capacitance. No shift was found. The amplitude was found to be around 190 mV slowly depending on the peaking time but not on the cumulated dose.

*2) Power Consumption:* We measured the whole power consumption of the IDeF-X ASIC on its mother board at each test run with 1 mA accuracy and we concluded that it was not affected by the irradiation.

*3) ENC Measurements:* We systematically measured the ENC versus peaking time characteristic of the channel #31 of the ASIC. Table II gives the values of the compensating current required to keep the channel operating without detector, for each test run.

TABLE II
IDEF-X HD COMPENSATING CURRENT

| RUN NUMBER | ACCUMULATED DOSE | PROGRAMMED CURRENT |
|---|---|---|
| 1 to 5 | 1 to 26 krad | 20 pA |
| 6 to 7 | 49 to 63 krad | 50 pA |
| 8 to 14 | 77 to 200 krad | 100 pA |

In IDeF-X HD, we used the same bonding pads as in IDeF-X V0 [10]. We already demonstrated that the leakage current in the protection diodes of the input pads increased with the dose. This is most likely the reason why we are constrained to increase the compensating current to keep the chip functional when the total ionizing dose increases.

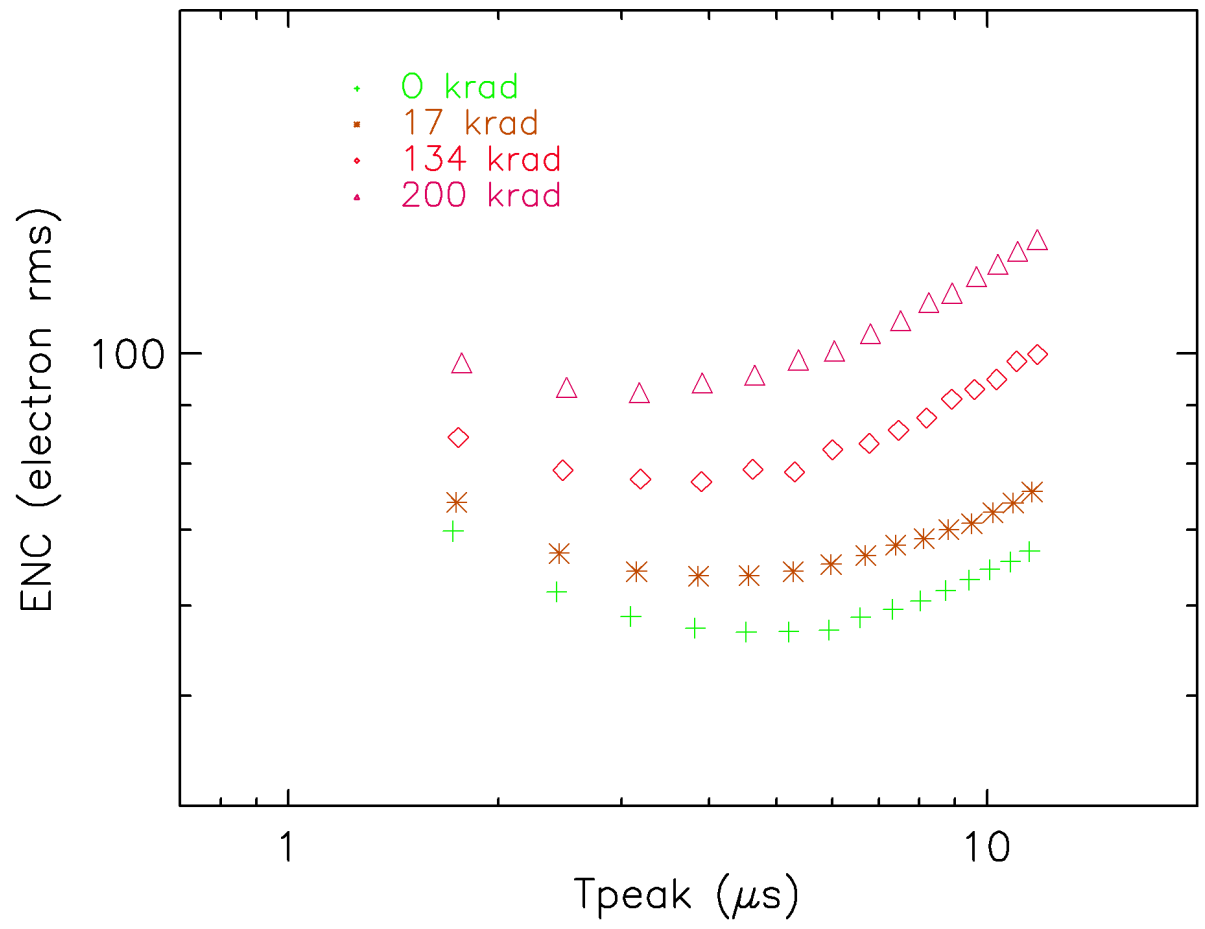

Fig. 19. ENC as a function of peak time at different accumulated dose.

Fig. 19 shows the ENC as a function of peak time for four different accumulated doses. As the dose increases, the parallel noise increases mainly due to the increase of the leakage current in the protection diodes and thus the optimal peak time also decreases. As shown in Fig. 20, the minimum ENC increases linearly with increasing dose. The slope is 0.17 $e^-$ rms/krad. Typical orbits and shielding for hard-X-ray astronomy require a total ionizing dose near 10 to 20 krad. More demanding orbits (to the Sun) may claim for a tolerance up to 100 krad. The excess noise due to the aging of the electronics will remain acceptable in any case since the noise remains permanently below 80 electrons.

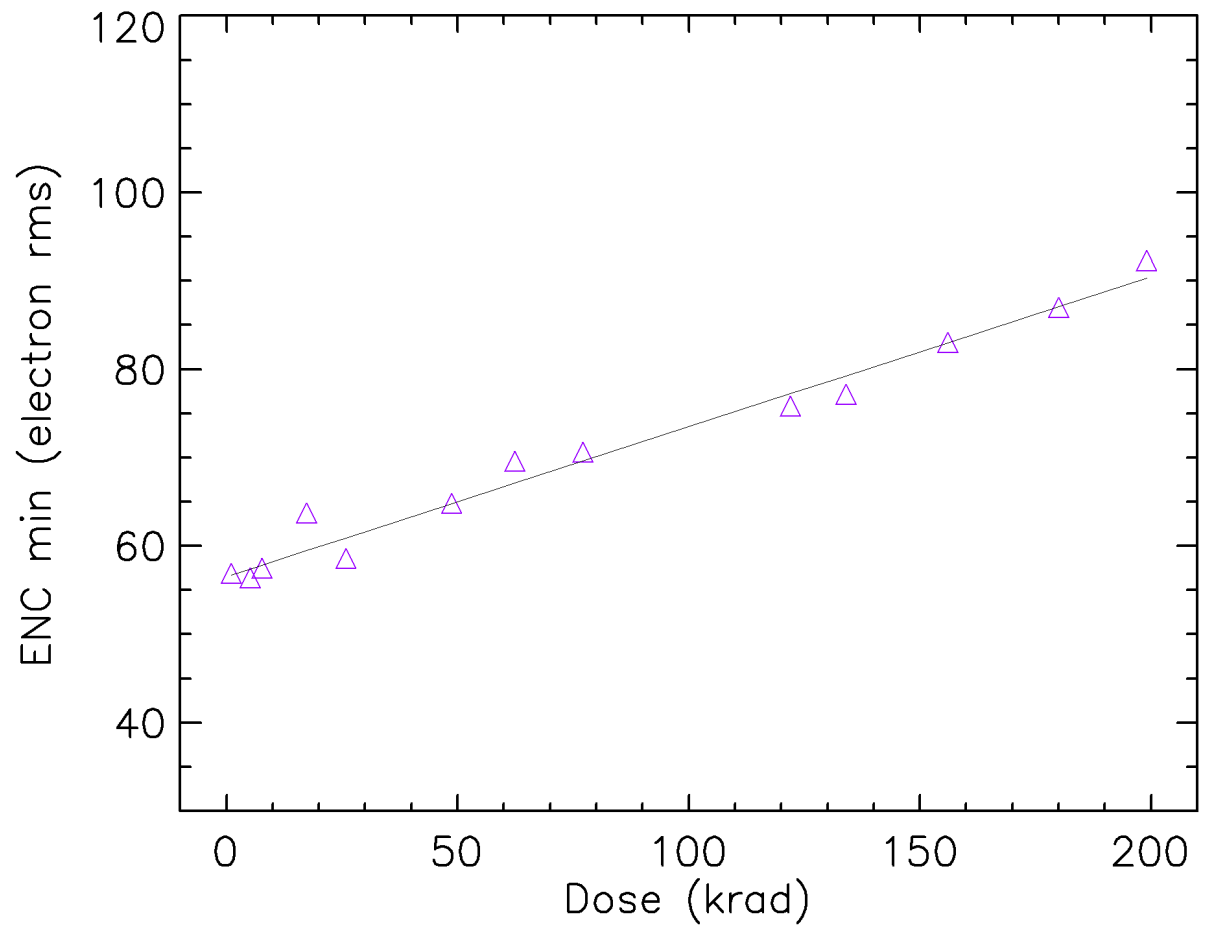

Fig. 20. Minimal ENC as a function of accumulated dose.

## 5.2 Heavy ion irradiation

### 5.2.1 Single Event Upsets (SEU)

To detect SEU, all the programming registers have been duplicated in the chip as in previous IDeF-X chip versions. [11]. IDeF-X HD integrates a SEU flag signal that can be used to reprogram the chip in case of SEU. SEU events are less critical than SEL as far as a simple software reset will efficiently restore the proper ASIC configuration.

### 5.2.2 Single Event Latchup (SEL)

Heavy ions induced SEL tests were performed at the Cyclotron of the University Catholique de Louvain (UCL) which is able to accelerate protons, alpha particles and heavy ions. The measurement setup is very much similar to the one described in [11]. We irradiated the ASIC with a set of different ions to cover a wide range of *Linear Energy Transfer* (LET) spectrum from a few MeV/(mg/cm$^2$) to higher than 100 MeV/(mg/cm$^2$). The chip under test was irradiated until an effective fluence of $10^6$ particles/cm$^2$ was reached with a flux of one thousand of particles/cm$^2$/s.

TABLE III
IDEF-X HD MAIN CHARACTERISTICS

| PARAMETER | VALUE |
|---|---|
| CHIP SIZE | 3500 µm × 5800 µm |
| TECHNOLOGY | AMS 0.35µm |
| POWER CONSUMPTION | 27mW (800 µW/channel) |
| GAIN | 50, 100, 150, 200 mV/fC |
| DYNAMIC RANGE | 36 fC (0.99 MeV for CdTe) at 50 mV/fC |
| DISCRIMINATION THRESHOLD | 90 $e^-$ to 3.6 k $e^-$ |
| PEAK TIMES (5%-100% OF SHAPED SIGNAL) | 0.73 µs to 10.73 µs (16 values) |
| ENC (FLOOR) | 33 $e^-$rms |
| ENC (200 KRAD) | 92 $e^-$ rms |
| LEAKAGE CURRENT TOLERANCE | Up to 4 nA |
| SEL LET | >110 MeV/(mg/cm$^2$) |

No single event latch-up occurred during test at all. It means that the LET threshold of the IDeF-X HD ASIC must be higher than 110 MeV/(mg/cm2). Since we used the same homemade SEL hardened digital library, we expected that the LET threshold of the digital power supply would be higher than 110 MeV/(mg/cm$^2$). However, as the whole analog circuitry of the ASIC is completely new compared to previous designs, we paid a peculiar attention to SEL in new blocks. During the design of the layout of the analog part of the chip, we respected the same rules as the one used in the digital part. Thus, we increased the tolerance of the chip to SEL to a similar LET threshold. The latch up immunity is a key characteristic of our circuits, which fulfils demanding requirements of space agencies (ESA and NASA).

# 6 Conclusion

The ASIC IDeF-X HD is a low power multi-channel integrated circuit. It is optimized for the readout of low capacitive (2-5 pF) and low dark current (~ 1 pA to few nA) Cd(Zn)Te mono-pixel detectors or pixel arrays for future X and gamma-ray astronomy space missions. The main parameters of the chip are summarized in Table III.
The gain of the channel can be set from 50 to 200 mV/fC. Thus, the dynamic range can be extended to 1 MeV for CdTe detectors. We successfully designed a low noise readout ASIC since the floor ENC was found to be 33 $e^-$ rms. Moreover, the power consumption is 800 µW per channel, four times lower than the power consumption of the previous chip versions. We succeeded in operating a CdTe Schottky detector with 1 keV FWHM energy resolution at 60 keV. The performances were confirmed in the high dynamic configuration experiencing a spectrum with $^{137}$Cs where the spectral performance was found to be 1.1 keV FWHM at low energy lines and 4.3 keV FWHM at 662 keV.
Eight of the IDeF-X HD ASIC have been used to build the new low power micro camera Caliste HD and we obtained an energy resolution of 666 eV at 60 keV at –4°C on the sum spectrum over the 256 pixels of the module. Eight of these new Caliste modules have been used to build the MACSI Camera, a 2048 fine pitch pixel array. Thanks to the drastic power reduction, we easily cool down the devices to low temperature where the noise performances are considered ideal to reach excellent energy resolution of the whole camera. Eventually, IDeF-X has been integrated into the Caliste-SO CdTe detector module populating the detector plane of the STIX instrument on board Solar Orbiter ESA mission [21]. The circuit is successfully operating in space since early 2020.